\documentclass[12pt,floats]{article}
\usepackage[]{umlaut,latexsym}
\usepackage[xdvi]{graphicx}
\usepackage{natbib}
\usepackage{amssymb}

\begin{document}
%%%%%%%%%%%%%%%%

\parskip 2mm
\renewcommand{\refname}{\normalsize \bf \em References}

%%%%%%%%%%%%%%%%%%%%%%%%%%%%%%%%%%%%%%%%%%%%%%%%%%%%%%%%%%%%%%%%%%%%%%%%%%%%
\title{\bf ISOLATED MODES AND PERCOLATION IN LATTICE DYNAMICS OF (Be,Zn)Se } 
%%%%%%%%%%%%%%%%%%%%%%%%%%%%%%%%%%%%%%%%%%%%%%%%%%%%%%%%%%%%%%%%%%%%%%%%%%%%
%     
\author{
      A.\ V.\ POSTNIKOV,$^a$\footnote{Corresponding author:
       Tel.: +49 541 969 2377,
       fax:  +49 541 969 2351,
       email: \mbox{apostnik@uos.de}}~
      O.\ PAGES,$^b$
      T.\ TITE,$^b$
      M.\ AJJOUN,$^b$ \\
      and J.\ HUGEL$^b$
\\*[0.2cm]
      $^a${\small \it Institute of Metal Physics,
      S. Kowalewskoj 18, Yekaterinburg 620219, Russia,} \\
      {\small \it and Universit\"at Osnabr\"uck -- Fachbereich Physik,
      D-49069 Osnabr\"uck, Germany} \\
      $^b${\small \it IPEM,  Universit\'e de Metz,  
      1 Bd Arago, F-57078 Metz cedex 3, France}
}               % author
\date{\small \it (Received \today)}
\maketitle
%%%%%%%%%%%%%%%%%%%%%%%%%%%%%%%%%%%%%%%%%%%%%%%%%%%%%%%%%

%%%%%%%%%%%%%%%%
\begin{abstract}
%%%%%%%%%%%%%%%%
%
A mixed II-VI semiconductor Zn$_{1-x}$Be$_x$Se possesses non-trivial vibration
properties, because its two constituent compounds, ZnSe and BeSe,
show very different degree of covalency and hence high elastic contrast.
An anomalous Be-Se vibration line has been observed mostly at
intermediate Be content in the Raman spectra of thin (Zn,Be)Se films.
In order to explain microscopic origins and the detailed composition
of these lines, a first-principles calculation of vibration frequencies
in a mixed crystal has been done, with frozen-phonon technique and
supercell setup within the density functional theory, by the
{\sc Siesta} method, which uses norm-conserving pseudopotentials 
and  strictly localized numerical basis functions. The calculations 
confirmed an earlier assumption that the anomalous Be-Se line appears 
due to the formation of continuous chains of a more rigid Be-rich 
pseudo-continuous phase formed within the more soft Zn-rich host region 
on crossing the Be-Se bond percolation threshold ($x\sim{0.19}$). 
Different local deformation in percolated and non-percolated regions affect 
interatomic elastic interactions and split corresponding vibration lines. 
Besides confirming the percolation model qualitatively, the calculation 
provides details about vibration patterns in different phonon modes. 
%  
%%%%%%%%%%%%%%
\end{abstract}
%%%%%%%%%%%%%%

\noindent {\it Keywords:}\/ Semiconductors, Density functional theory, 
Phonons, Percolation, Isolated modes

%%%%%%%%%%%%%%%%%%%%%%%%%%
\section*{1. INTRODUCTION}
%%%%%%%%%%%%%%%%%%%%%%%%%%

Chalcogenides of Be, characterized by a strong covalent character of their 
chemical bonding, are outstanding among the II-VI semiconductors 
which are otherwise largely ionic systems. BeSe makes solid solution
with ZnSe in the whole composition region. Beyond providing the possibility
to tune the fundamental band gap and dielectric function in broad limits
\citep{PRB59-10071}, the coexistence of
strongly ionic and strongly covalent bonds in the same crystal makes
its elastic and vibrational properties interesting.
The frequencies of optical modes 
in pure ZnSe and BeSe are far separated. 
This remains basically true in the alloy \citep{APL77-519}. The remarkable
point is that an additional Be-Se mode has been detected, mostly in the
intermediate concentration range, 20 to 80 at.\% Be. 
This effect has been tentatively attributed by \citet{PRB65-035213}
to the onset of percolation on the Be sites, i.e., creation of
infinite --(Be--Se)-- chains in the predominantly ZnSe crystal.

In the attempts to understand and describe the lattice dynamics 
of mixed crystals in terms of simple models, an obvious difficulty is 
that interaction parameters (force constants) between two given atoms depend
on their local environment in a not straightforward way. The elastic
couplings are affected by interatomic distances which may differ in an alloy;
moreover the combinatorial aspect (how many neighbours of one or another
type) may play an unknown role. Therefore it could be difficult to agree
on a simple and yet realistic model.
First-principles calculations of lattice vibrations, either by linear response
\citep{PRB48-3156,PRB50-13035,RMP73-515} or frozen-phonon 
\citep{PRL69-2799} schemes, work well for perfect crystals
\citep{PRB60-15511}
or ordered mixed phases \citep{CMS13-232,RMP73-515} 
but face problems in case of substitutional disorder. 
In this latter case, the treatment of representative
supercells remains a virtually unique alternative, with the obvious
shortcoming that the selection of such supercells is limited, 
and the calculation technically demanding.
In the present contribution, we calculate electronic structure,
force constants and phonon frequencies in the (Zn,Be)Se mixed system
using the first-principles method, and calculation code, 
{\sc Siesta} \citep{JPCM14-2745}. The calculations 
were practically done for Zn$_3$BeSe$_4$, ZnBe$_3$Se$_4$ and 
Zn$_{26}$Be$_6$Se$_{32}$ supercells, the latter simulating both percolating 
and ``isolated'' Be ions in the lattice.

The additional vibrational modes, whose appearance in the calculation
could be traced back to the onset of percolated Be chains, closely resemble 
those observed earlier in experimental spectra by \citet{PSSB229-25} 
and outlined briefly in Sec.~2. This supports earlier assumptions about 
the role of percolation in the formation of split-off vibration lines. 
The results of calculation, which were organized as described in Sec.~3,
allow to study the composition of each vibration mode separately, and 
to explain the underlying microscopics, in terms of variation of 
interatomic distances. The discussion of results is given in Sec.~4.

%%%%%%%%%%%%%%%%%%%%%%%%%%%%%%%%%%%%
\section*{2. EXPERIMENTAL SITUATION}
%%%%%%%%%%%%%%%%%%%%%%%%%%%%%%%%%%%%

\begin{figure}[b!] 
\begin{center}
\includegraphics[angle=0,width=11cm]{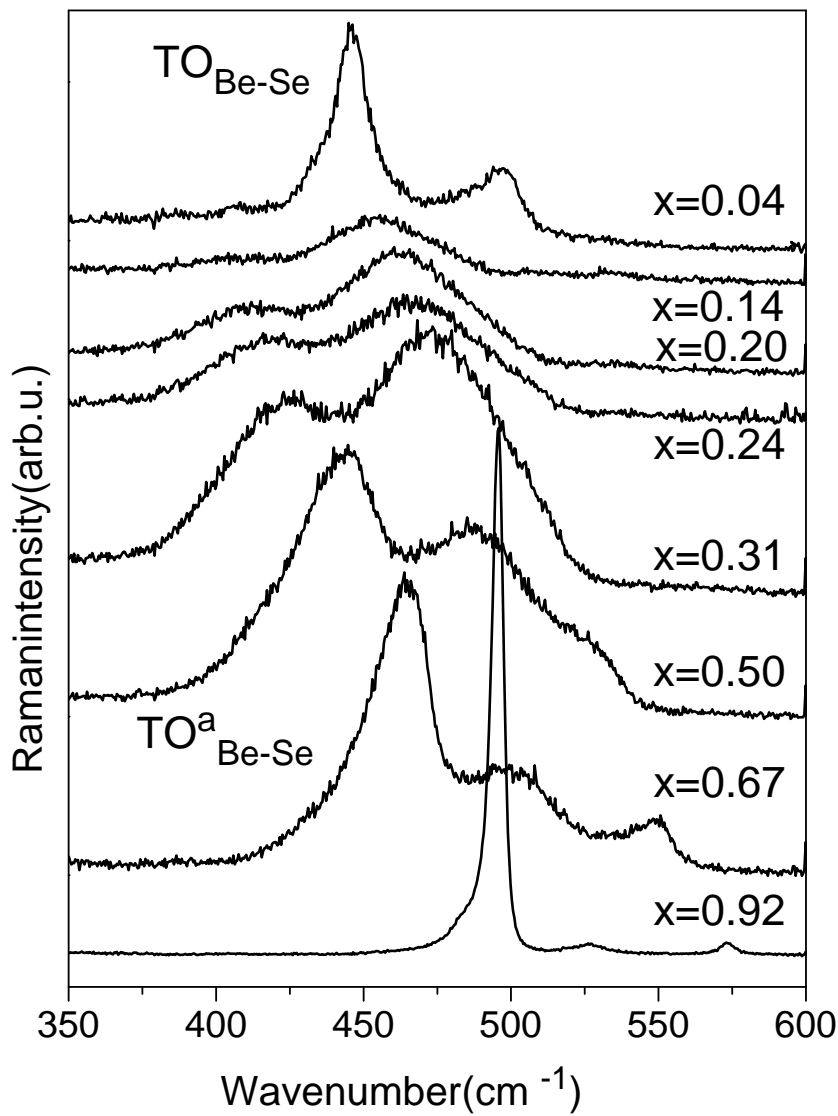}
\end{center}
\caption {\small
Raman spectra of Zn$_{1-x}$Be$_x$Se films grown on the GaAs substrate,
measured in the backscattering geometry along the (110) edge axis.
The ``regular'' and ``anomalous''
(labeled TO$^{\,\mbox{\footnotesize a}}_{\,\mbox{\footnotesize Be-Se}}$)
lines coexist in the concentration range 20 to 80 at.\% of Be.
See \citet{PRB65-035213} for more detailed discussion of the experiment.
\label{fig:exp}
}
\end{figure}

The vibrations spectra of (Zn,Be)Se alloys have been detected  
by Raman measurements in the backscattering geometry from thin films grown
by molecular beam epitaxy on (001) GaAs. 
The experimental setup, obtained spectra and their discussion have been
in detail described earlier \citep{APL77-519,PRB65-035213,JAP91-9187}.
A typical selection of spectra,
in dependence on concentrations, is shown in Fig.~\ref{fig:exp}. 
For the sake of our subsequent discussion, we emphasize the region
of transverse optical (TO) Be-Se modes, in the range 400--500 cm$^{-1}$.
The Zn--Se mode, around 200 cm$^{-1}$, is useless because of strong
Fano interference with a disorder-activated acoustical
continuum \citep{PRB65-035213}. Incidentally the large frequency gap
between the Be--Se and Zn--Se nodes is basically attributed to the large
difference in the reduced mass of two bonds, in the ratio 1:4.
The increase in the Be concentration shifts the BeSe-like
TO line upwards. From $\sim$20\% of Be on, an additional (``anomalous'') peak
is formed at $\sim$70 cm$^{-1}$ on the low-frequency side from the main one.
Two peaks coexist through the concentration range $\sim$20 to $\sim$80\%
of Be, both displacing towards higher frequencies and approximately
maintaining their separation. The intensity of the anomalous peak grows
at the expense of the normal one, until finally (on the Be-rich side)
the anomalous peak remains the only one, and ends up as the narrow
TO line in the limit of pure BeSe. 

The discussion in previous works \citep{PRB65-035213,PSSB229-25} 
emphasized that such behaviour cannot be accounted for by segregation 
of crystal into BeSe-rich and ZnSe-rich fractions. 
The proposed explanation of the anomaly was the following. 
Above the Be--Se bond percolation threshold ($\sim$19 at.\% Be),
a pseudo-infinite wall-to-wall chain of Be--Se bonds is formed spontaneously
throughout the alloy. Based on simple local strain considerations,
the bond length within the chain was presumed to be larger than the bond
length of the Be--Se bonds isolated in the ZnSe-like host matrix, with
concomitant impact upon the phonon frequencies. However, microscopic
justification of this picture was so far lacking. Our present aim is
to construct a representative structure model,
compare the vibration properties following from it with experimental
observations, and in case of agreement to analyze microscopical
origins of the observed anomaly.

%%%%%%%%%%%%%%%%%%%%%%%%%%%%%%%%%%%%%%%%%%
\section*{3. CALCULATION METHOD AND SETUP}
\label{sec:calc}
%%%%%%%%%%%%%%%%%%%%%%%%%%%%%%%%%%%%%%%%%%

First-principles calculations have been performed within the general
scope of the density functional theory
\citep[see, e.g.,][for a review]{RMP71-1253},
using the calculation method {\sc Siesta} \citep{JPCM14-2745}
with its compact basis of strictly confined atom-centered functions
\citep{JPCM8-3859,PRB64-235111}. Double-$\zeta$ basis functions with
polarization orbitals were used for all valence states -- Be $2s$ and $2p$, 
along with $3d$, $4s$ and $4p$ for Zn and Se. Note that we attributed
the semicore Se$3d$ states to the valence band. 
The norm-conserving pseudopotentials
have been generated after \citet{PRB43-1993}. The results discussed below
have been obtained using the local density approximation for the
exchange-correlation. 

\begin{figure}[t!] % Fig. 2
\begin{center}
\includegraphics[angle=0,width=13cm]{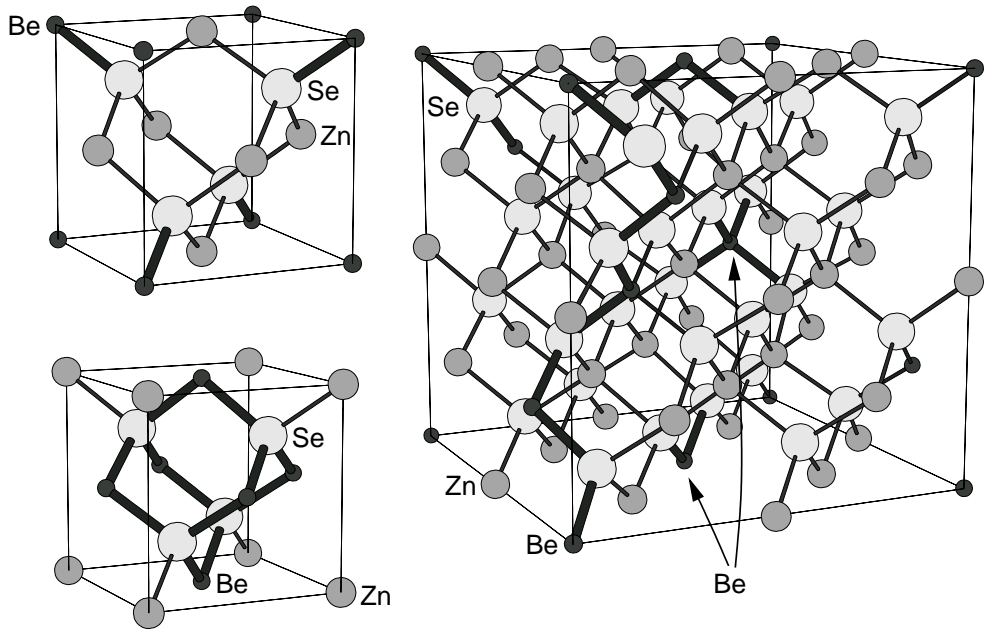}
\end{center}
\caption {\small
BeZn$_3$Se$_4$, Be$_3$ZnSe$_4$ and Be$_6$Zn$_{26}$Se$_{32}$ supercells
used in the calculation. More covalent Be--Se bonds are shown
by thick lines. The large supercell contains a continuous
(--Be--Se--) chain and two isolated Be atoms (indicated by arrows),
who have only Zn atoms as next-nearest neighbours.
\label{fig:scells}
}
\end{figure}

The calculations have been done for three different cubic supercells,
shown in Fig.~\ref{fig:scells}. The smaller ones
included four formula units in the zincblende structure.
BeZn$_3$Se$_4$, with the nominal Be concentration (25 at.\%) already
beyond the percolation threshold, contained nevertheless only
isolated (and artificially ordered) Be impurities, whose second-nearest
neighbours are Zn atoms only. The ``symmetric'' case of 75 at.\% Be
(Be$_3$ZnSe$_4$) contains, on the contrary, isolated Zn atoms and
fully connected Be-Se network.
Moreover, a larger supercell Be$_6$Zn$_{26}$Se$_{32}$
simulated the onset of percolation on the Be sublattice, i.e.,
the formation of continuous (--Be--Se--) chains, along with the presence
of isolated Be substitutions. The Be concentration in this case is just
above the percolation threshold on the zincblende lattice.
In all cases, the calculated forces and strain were used for 
relaxing the lattice vectors and internal coordinates
by the conjugated gradient algorithm, without any symmetry constraints 
imposed. This allowed us to discuss the variation of the lattice parameters
and interatomic distances in their dependence on concentration and
local environment.
The vibration frequencies have been calculated in the frozen phonon approach,
sequentially introducing finite displacements of atoms in the supercell
by 0.03 Bohr along three Cartesian directions from their equilibrium
positions, and analyzing the forces induced on all atoms.
In a single cell, this technique gives access to the zone-center TO
phonon only, but in a larger supercell the analysis of phonon eigenvectors
allows to make some conclusions about the phonon dispersion.

\begin{table}[t!]
\caption{\small
Equilibrium lattice constant $a$, bulk modulus $B$, and
frequency of the $\Gamma$-TO phonon $\omega$ for pure ZnSe and BeSe
from the present calculations (WIEN97 and {\sc Siesta}), previous
calculations, and experiment.
}
\begin{center}
\begin{tabular*}{\textwidth}{l@{\extracolsep\fill}ccccccc}
\hline
 \rule[-2mm]{0mm}{7mm}
 & \multicolumn{3}{c}{ZnSe} && \multicolumn{3}{c}{BeSe} \\
 \cline{2-4} \cline{6-8}
 Method & $a$ ({\AA}) & $B$ (Kbar) & $\omega$ (cm$^{-1}$) &&
 \rule[-2mm]{0mm}{7mm}
          $a$ ({\AA}) & $B$ (Kbar) & $\omega$ (cm$^{-1}$) \\
\hline \hline
WIEN97       & 5.571 &   727    & 162     && 5.087     & 831 & 456 \\
{\sc Siesta} & 5.590 &   778    & 200     && 5.114     & 965 & 429 \\
exp.         & 5.668 & 624--647 & 207$^a$ && 5.137$^b$ & 920 & 501$^c$ \\
other calc.  &       & 750$^d$  & 216$^e$ &  \\
\hline
\end{tabular*} \\*[4mm]
\end{center}
\footnotesize
$^a$\citet{CrRT38-359};
$^b$\citet{PRB52-7058};
$^c$\citet{APL77-519}; 
$^d$\citet{PRB32-7988};
$^e$\citet{PhL36A-376} 
\normalsize
\end{table}

As benchmarks for testing the quality of the pseudopotentials
used and of the basis sets, we calculated the equilibrium lattice parameter
and the $\Gamma$-TO frequency in pure ZnSe and BeSe. The calculation results
are shown in Table I along with the experimental data and the results
of another, all-electron calculation, which has been performed
using an accurate all-electron full-potential linearized augmented 
plane wave method WIEN97 \citep{wien97}.
The difference between the {\sc Siesta} results and the experiment
reflects the shortcomings of the local density approximation, and moreover
technical limitations of the pseudopotential generation and the choice
of basis functions. A similarly more faithful estimation of some
results with {\sc Siesta}, as compared to WIEN97, in view of numerical 
superiority of the latter method, is probably due to some error cancellation.
Anyway, Table I gives an idea of the level of accuracy of our present
calculation approach. It is obvious already from comparison the
bulk moduli that ZnSe is much softer than BeSe. The same trend holds
for the force constants, and mainly accounts for the difference in 
the $\Gamma$-TO phonon frequencies. The difference in Be and Zn masses 
further enlarges this difference.

%%%%%%%%%%%%%%%%%%%%%%%%%%%%%%%%%%%%%%%%%%%%%%%%% 
\section*{4. RESULTS AND DISCUSSION}
\label{sec:results}
%%%%%%%%%%%%%%%%%%%%%%%%%%%%%%%%%%%%%%%%%%%%%%%%%

\begin{figure}[t!] % Fig. 3
\begin{center}
\includegraphics[angle=0,width=11cm]{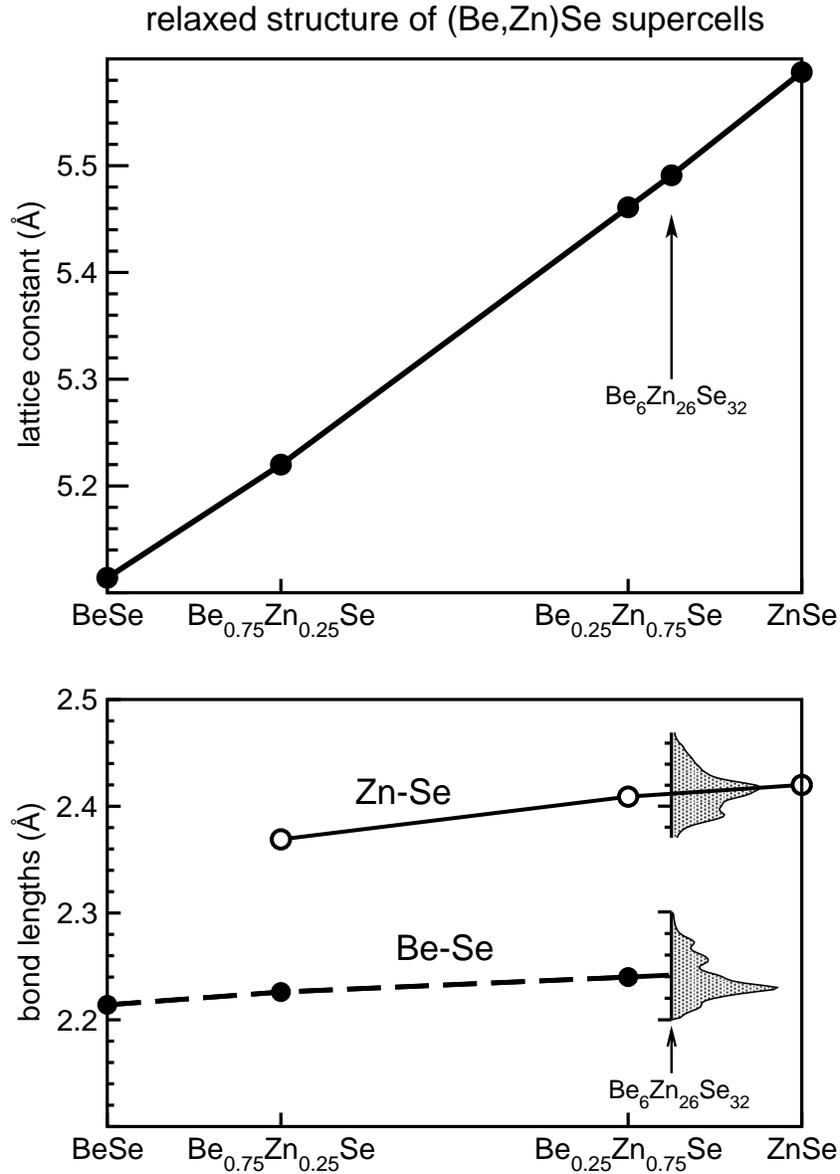}
\end{center}
\caption {\small
Mean lattice constant (top panel) and nearest-neighbor bond lengths (bottom
panel), according to calculations on two pure end systems and
three mixed-composition supercells. In the Be$_6$Zn$_{26}$Se$_{32}$ supercell,
the distribution of Be--Se and Zn--Se bond lengths is shown.
\label{fig:bonds1}
}
\end{figure}

Fig.~\ref{fig:bonds1} shows the calculated equilibrium lattice constant
(derived from the volume of relaxed supercells) and nearest-neighbour 
distances in pure materials and three mixed-composition supercells.
One can see that the (Zn,Be)Se lattice parameter quite accurately
follows the Vegard's law. At the same time, the Be--Se and Zn--Se bond
lengths, which differ by about 9\% in the parent compounds,
tend to remain diverse, and almost independent
on concentration. It means that the accommodation of short Be--Se and
long Zn--Se bonds in the lattice of a mixed crystal may lead to
local strains and frustrations. In fact, in the large supercell
one finds a distribution of bond lengths of one and another type,
shown as continuous curves in the bottom panel of Fig.~\ref{fig:bonds1}.%
\footnote{%
Obviously the number of nearest-neighbour bonds in the 64-atom supercell
is finite (=128); an artificial smearing of the bond distribution
function is introduced for better visibility.}
A further analysis of these different bond lengths shows that isolated
(non-chain) Be atoms can easily induce a local and ``isotropic''
lattice contraction, binding all their four Se neighbours at a distance 
of 2.23 {\AA}, i.e. at nearly the nominal bond length of pure BeSe.
About the same, or even shorter (2.21 to 2.23 {\AA}), bond length 
gets stabilized between Be atoms in the chains and their off-chain 
Se neighbours. One finds finally that the Be--Se bonds
\emph{along the chains} are the most difficult to shorten, 
and exhibit a variation of lengths 2.24 -- 2.28 {\AA}. This can be
understood because the chain is infinite, and its period is
essentially fixed by the mean lattice constant of a crystal with 
a predominant Zn concentration. The only possibility to shorten the chain
links is at the price of increasing the Se--Be--Se angles, making 
a zigzag chain more straight. This increases average distances
between in-chain Se atoms and their Zn neighbours beyond the comfortable
ZnSe bond length. 

\begin{figure}[t!] % Fig. 4
\begin{center}
\includegraphics[angle=0,width=11.5cm]{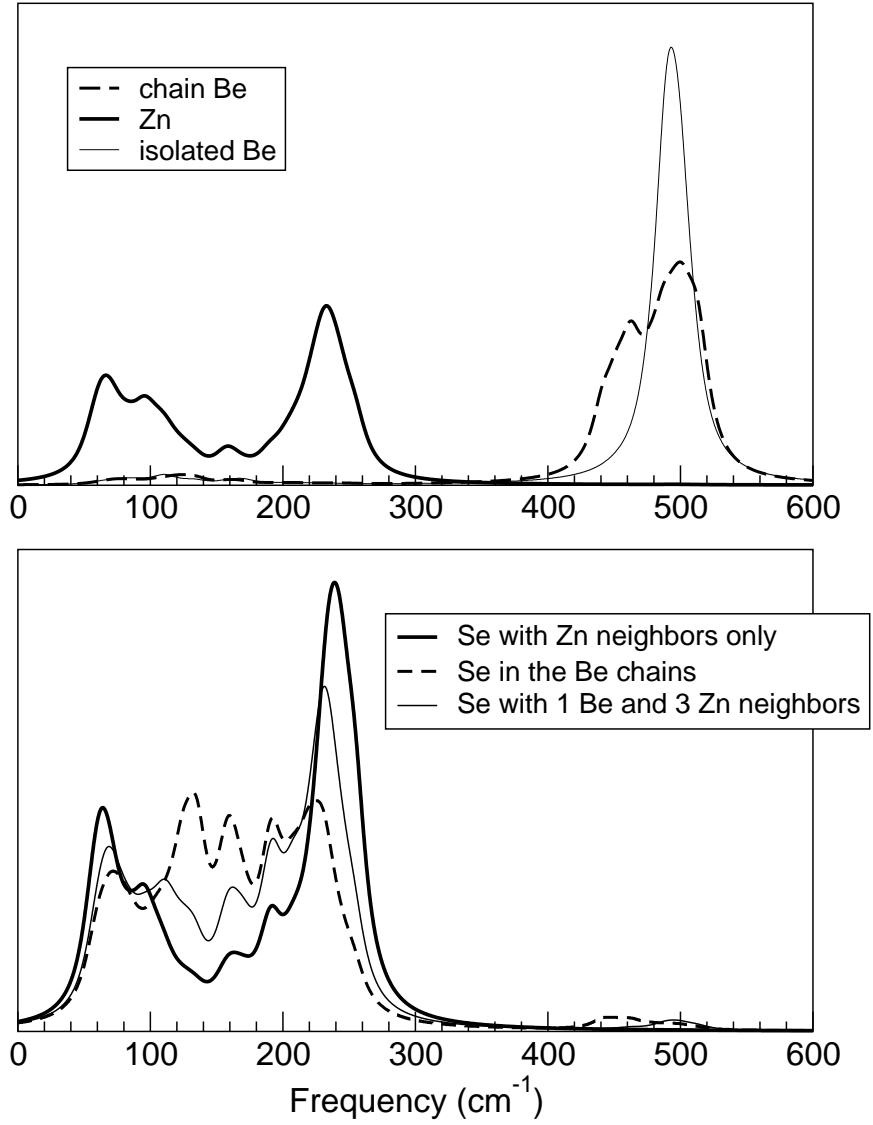}
\end{center}
\caption {\small
Phonon densities of states for different categories of atoms
represented in the Be$_6$Zn$_{26}$Se$_{32}$ supercell.
Top panel: Be and Zn contributions, bottom panel: Se contributions.
Three acoustic modes at $\omega$=0 are removed, and the broadening of discrete
vibration lines is introduced. 
\label{fig:PhDOS}
}
\end{figure}

The discussed differences in interatomic distances, and their stretching
beyond, or below, the natural bond lengths has an effect on the calculated
force constants. A diagonalisation of the dynamical matrix yields frequencies
and eigenvectors of all 192 zone-center vibration modes of the supercell, 
including three acoustic modes (the closeness of whose frequencies to zero
is a good test for the numerical accuracy of the calculated forces).
Fig.~\ref{fig:PhDOS} shows phonon densities of states, constructed as
%
%%%%%%%%%%%%%%%%%
\begin{eqnarray*}
%%%%%%%%%%%%%%%%%
%
n_{\{\alpha\}}(\omega) = \sum_{\{\alpha\}} \sum_i
\frac{\Delta}{\pi}\,
\frac{|\mathbf{A}_{\alpha}(\omega_i)|^2}{\Delta^2 + (\omega-\omega_i)^2}\,,
%
%%%%%%%%%%%%%%%%
\end{eqnarray*}
%%%%%%%%%%%%%%%%
where $\omega_i$ are (discrete) vibration frequencies in the supercell
(leaving aside the three zero-frequency acoustic modes),
$\mathbf{A}_{\alpha}$ are elements of the eigenvector of the corresponding
mode, and the summation is done over $\{\alpha\}$, a selection of atoms
in a certain structure category. A broadening parameter $\Delta$=10 cm$^{-1}$
was introduced for better visibility.

The top panel of Fig.~\ref{fig:PhDOS} shows the contributions of Zn and Be
atoms, which are well separated in the frequency domain. The lower part
(up to $\sim$300 cm$^{-1}$) resembles the phonon density of states 
in ZnSe \citep[see, e.g.,][]{PhL36A-376}. The high-frequency part
represents a Be--Se vibration; its fine structure reveals the differences
in the force constants which couple different Be--Se pairs.
The vibration modes involving the \emph{isolated} Be are densely
situated around 490 cm$^{-1}$. On the contrary, the vibration bands involving
Be atoms in the chains split into a softer and a harder components.
A comparison with the phonon DOS projected on Se atoms (Fig.~\ref{fig:PhDOS},
bottom panel) shows that the softer one affects almost exclusively
the Se atoms in the chains (i.e those with two Be and two Zn neighbours),
whereas the harder one involves also the peripheric Se atoms,
which build just one bond to Be in the chains. This separation immediately 
follows from the distinction between extended Be--Se bonds along the chain,
with correspondingly weakened force constants, and ``natural''
Be--Se bond lengths to off-the-chain Se atoms, or around isolated Be
impurities. This observation clearly indicates that the splitting off
of a softer ``additional'' line from the main TO Be--Se peak in the
measured spectra (Fig.~\ref{fig:exp}) is due to the formation of 
continuous (--Be--Se--) chains, where the Be--Se bond length is essentially
fixed by the predominantly ZnSe bulk, therefore extended over the
``natural'' Be--Se distance, and results in weakened force constant.
The drift of the ``normal'' and ``additional'' lines with concentration,
seen in Fig.~\ref{fig:exp}, is not directly addressed by our simulation 
on a single ``percolated'' supercell, but it can be easily understood
from the general trend of the average lattice constant. A gradual
shortening of Be--Se distances with the increase in Be concentration
hardens \emph{all} vibration modes.
The concept of percolation on the Be sublattice also makes clear why
the ``additional'' and not the ``normal'' mode transforms into the single 
TO mode in the limit of high Be concentration (see Fig.~\ref{fig:exp}):
the ``additional'' mode was that involving chained (--Be--Se--) structures,
which at high Be concentration become omnipresent.

\begin{figure}[b!] 
\begin{center}
\includegraphics[angle=0,width=11.5cm]{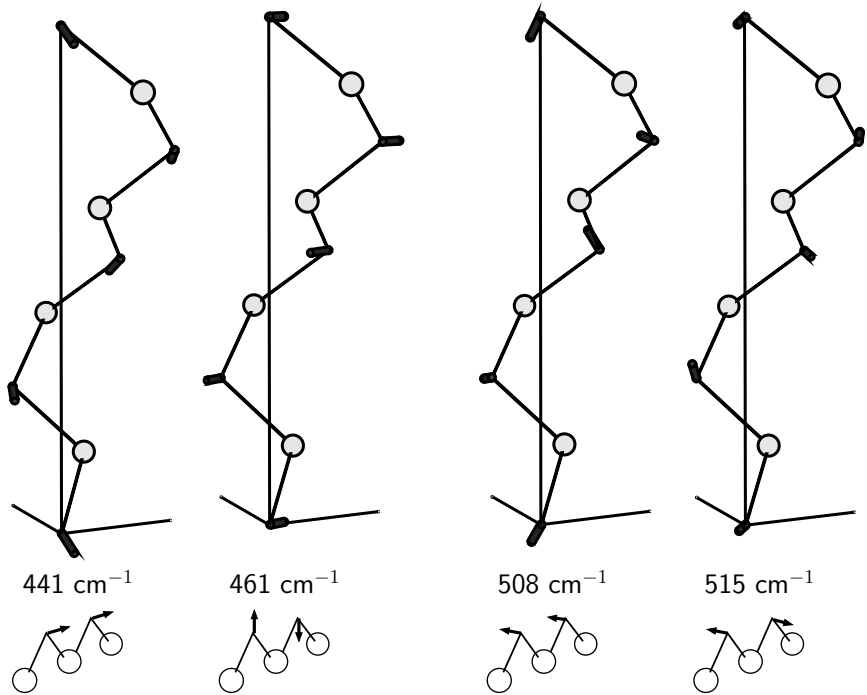}
\end{center}
\caption {\small
Snapshots of four selected vibration modes involving Be atoms
in the chains. The displacements of Se atoms are negligible
in the chosen scale. For each mode, a simplified schema of
its characteristic displacement is shown below.
\label{fig:modes}
}
\end{figure}

In order to get a better idea of vibration patterns within ``normal'' and
``additional'' lines, we shown in Fig.~\ref{fig:modes} the snapshots of
four vibration modes. Whereas the most of vibration modes involve many
atoms in the supercell in a complicated way, these four allow more or less
easy understanding of their underlying displacement. Such simplified
schemas are shown below each mode, neglecting the bending of the zigzag
chain. The most general observation is that the movement of Be atom
in the common plane with two Se neighbours occurs at lower frequency
than out-of-plane vibrations. This is however a quite simplifying
statement; one should consider that the off-the-chain Se atoms also
participate in these movements, up to different extent in harder
(``normal'') and sifter (``additional'') modes, as has been mentioned above.

We leave for a subsequent more detailed discussion elsewhere 
the analysis of vibrations in smaller supercells, which we used here only
as a reference to the variation of lattice parameters. Moreover, the analysis
of vibration modes with respect to their wave vector is important and
will be reported separately, as only $\mathbf{q}$=0 vibrations contribute
to the Raman spectra.

%%%%%%%%%%%%%%%%%%%%%
\section*{5. SUMMARY}
%%%%%%%%%%%%%%%%%%%%%

By performing a straightforward calculation of vibration frequencies
in a large supercell simulating (Zn,Be)Se mixed system,
we were able to explain the onset of ``anomalous'' vibration mode
in the frequency range close to the $\Gamma$-TO line of BeSe. This mode
appears due to a softening of corresponding force constants in expanded
Be--Se bonds, that occurs as continuous Be--Se chains start to traverse
the Be$_x$Zn$_{1-x}$Se crystal on reaching the Be percolation limit
of $x$=19 at.\%. We found remarkable variations of the Be--Se bond length in
a (Zn,Be)Se mixed crystal with the Zn predominance. The Be--Se
distances shorten most efficiently between Be ions in the percolating
chains and their neighbouring Se ions \emph{not} belonging to the
chains. On the contrary, the Be--Se distances within the chains
are essentially fixed by the ZnSe bulk, i.e. they are over-stretched,
as compared to relaxed BeSe crystal. The Be--Se bond length for
a Be ion being an isolated substitutional impurity in the ZnSe matrix
is intermediate between two previously mentioned cases. As a consequence,
one finds distinct Be--Se force constants for these three cases, 
and hence the splittings of corresponding vibration modes.

%%%%%%%%%%%%%%
\end{document}